# Dissertations Repository System Using Context Module

Ali K.Hmood, M.A.Zaidan, Hamdan.O.Alanazi, Rami Alnaqeib, Yahya Al-Nabhani

**Abstract:-** Without a doubt, the electronic learning makes education quite flexible. Nowadays, all organizations and institutions are trying to avoid Monotony and the delay and inertia. As well the universities should be improving their systems continually to achieve success. Whereas, the students need to access the dissertations in the library. In this paper we will present Dissertations Repository System Using Context Module to allow the students to benefit the dissertations which is in the library flexibly.

——————————— ◆ ———————————

## 1. INTRODUCTION

The objective of this system is to develop a Dissertations Repository System (DRS). We will use the context diagram to show the environment of the system. Descriptions are provided for the important characteristics of the DRS system [1]. The system will allow students access to the library materials through an online login system [2]. In creating and hosting digital collections, the goal of this paper is to implement and build a system to help the students to get any dissertation easily and provide transparency, Flexibility, Reliability, Usability, and Integrity [3].

## 2. SYSTEM MODULE

System modeling helps the analyst to understand the functionality of the system and models are used to communicate with customers [4].

### 2.1 Context Diagram

A context diagram is a data flow diagram, with only one massive central process that subsumes everything inside the scope of the system. It shows how the system will receive and send data flows to the external entities involved. Here's a theoretical example [5]. See Fig 1

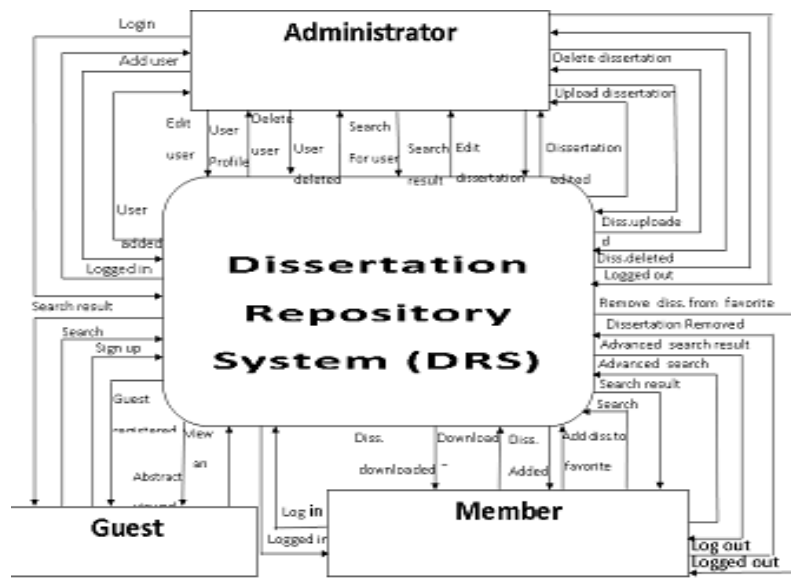

Fig 1. Context Diagram

## 2.2 System Environment

Database: collection of information organized in such a way that a computer program can quickly select desired pieces of data [6].

Repository: It is a place where all information for our system stored.

In our system the place is the database and the stored members profile, dissertation information and dissertation files.

- The system contains information about: Dissertations of students, Dissertation's Authors.
- The system using PHP (Hypertext Preprocessor) programming language to build a content of the website and MYSQL to manage the database of the system.
- All Dissertations will be uploaded to Internet server.
- The users must have ISP (Internet service provider) to access the system.
- The Administrator can manage the system online.
- A user of system may access the repository system and the search engine through a user interface on the Website.
- A user can access the system using any Internet browser such as: Internet Explorer, Mozilla Firefox, and Opera.

More than one user can access the system at the same time.

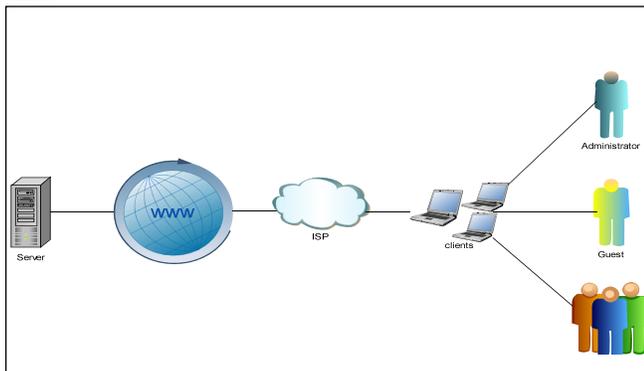

Fig 2. A Process of Showing the Delivery System to Member and Control by an Administrator.

## 2.3 Functional Requirement Specification

Listed below are the functional requirements specifications [7]:

- Function: Login

Definition: For the users to be able to use this system, they have to enter username and password which they have created before and been saved in the database in the Login page. The user might be a normal user or an Admin.
Inputs: Username and password.
Outputs: The system will state whether inputs are correct or not.
Pre conditions: The user should have signed up in the system and have a valid username and password.
Post conditions: The user will enter the main page of him/her self.

- Function: Sign Up

Definition: When a new student registers in the faculty and gain his/her matrix number, he /she can then sign up in the system using the given number.
Inputs: Matrix number, valid username, password, e-mail address, etc.
Outputs: The system will state whether the matrix number exists in database or not and also will check if the username is available or not.
Pre conditions: There is no prerequisite for entering this page of the system.
Post conditions: When the process finishes successfully, the user will be guided to the main page where he/she can go to Login page and use the facilities.

- Function: Add User

Definition: when a new student register in the faculty, information about him/her will be handed to the Admin to upload it in the database for further usage such as sign up.
Inputs: Matrix number, user's name, Master/PhD, etc.
Outputs: The system will state that the information has been uploaded successfully in the database.
Pre conditions: Being a valid Admin and clicking the "Add user" icon in the Admin main page.
Post conditions: When the process finishes successfully, Admin will be guided to the main page where he/she can use other facilities.

- Function: Delete User

Definition: When a student is not supposed to use the system, once it has been decided by the management, the Admin will be informed and he/she can delete this user from the database.
Inputs: Matrix number or user's name.
Outputs: The system will state that specific user has been deleted successfully from the database.
Pre conditions: Must be a valid Admin and can search the user's profile to display the search result.
Post conditions: After deleting the user, Admin will be guided to the search results again, but this time the number of results is 1 less.

- Function: Edit User

Definition: This function gives the authority to the Admin for editing user's profile. A valid Admin can change student's name, matrix number, etc in order to input or correct any information.
Inputs: Matrix number or user's name.
Outputs: The system will state that specific user's profile is saved successfully in the database.
Pre conditions: Must be a valid Admin and can search the user's profile to display the search result.
Post conditions: After editing the user's profile, Admin will be guided to the search results again.

- Function: Search User

Definition: This function is used when Admin wants to find specific user to do some functions such as delete, edit, etc.
Inputs: User's matrix number and/or user's name.
Outputs: The system will come up with appropriate search results page.
Pre conditions: Must be a valid Admin and can click on "User Process" button.
Post conditions: Admin can do some changes in database according to the search results such as delete a user, edit user's profile, etc.

- Function: Logout

Definition: This function is used when a logged in user finishes his/her job and wants to be logged out so that no one can abuse his username.
Inputs: N/A
Outputs: The system will state the user has been logged out successfully.
Pre conditions: The user should have logged into the system.
Post conditions: The user will enter the main page of the system (Index).

- Function: Search

Definition: The most important function of the system is finding dissertations. To do so any one can use the normal search function, even a guest. This function uses a search engine to explore the database.
Inputs: keyword(s).
Outputs: The search results.
Pre conditions: N/A.
Post conditions: The search results page will appear with some functions such as view and print results, but if the user has been logged in, he/she can also download any of the dissertations.

- Function: Advance search

Definition: This function is very similar to the normal search function, but with much more options and fields such as author, date, topic, etc.
Inputs: keyword(s) in appropriate fields.
Outputs: The search results.
Pre conditions: Being a logged in user.
Post conditions: The search results page will appear with all possible functions such as view, print results, download and add to favourite.

- Function: Remove favourite

Definition: Any logged in user has a favourite page that can be included as his/her favourite dissertations so that he/she does not have to search again to find them. There is remove option in this page which clears one or more dissertations from the list.
Inputs: Selecting dissertation's name.
Outputs: The system will state the dissertation has been removed successfully.
Pre conditions: Must be a logged in user to click the favourite button.
Post conditions: The user will still be in his/her favourite page without selecting any dissertations.

- Function: Add to favourite

Definition: Any logged in user has a favourite page, which is included as his/her favourite dissertations so that he/she does not have to search again to find them. After finding a dissertation with search process, user can add it to his/her favourite list by clicking the appropriate button.
Inputs: Selecting dissertation's name.
Outputs: The system will state the dissertation has been added to the list successfully.
Pre conditions: Must be a logged in user to perform a search results.
Post conditions: The user will still be in his/her search results page to apply more functions.

- Function: Upload dissertation

Definition: When any dissertation is approved by the lecturers and dean, it can be uploaded in the database by Admin so that other students can use it.
Inputs: Dissertation's appropriate file.
Outputs: The system will state the dissertation has been uploaded successfully.
Pre conditions: Must be a valid Admin to click "Upload" button.
Post conditions: Admin will be guided to his/her main page after the process finishes successfully.

- Function: Delete dissertation

Definition: It seems like deleting a dissertation is unnecessary because it is unlike that an approved dissertation will be deleted from the system. However, we include this option in the system in order for the management to perform the delete function of any dissertation if they found it necessary.

Inputs: Dissertation's name.
Outputs: The system will state the dissertation has been deleted from the database successfully.
Pre conditions: Must be a valid Admin and have the search results page for the appropriate dissertation to be displayed.
Post conditions: Admin will be guided to his/her search results page after the deleting process completed successfully, but without that dissertation in the list.

- Function: Edit dissertation's profile

Definition: Sometimes it is necessary to edit some information about the dissertation due to keying information incorrectly. Only Admin is allowed to edit dissertation's profile.
Inputs: Correct information for the profile.
Outputs: The system will state the dissertation's profile has been saved in the database successfully.
Pre conditions: Must be a valid Admin and show the search results page for the appropriate dissertation and click "Edit dissertation's profile" button.
Post conditions: Admin will be guided to his/her search results page after the process completed successfully.

## 2.4 Non-Functional Requirements

There are some constraints on the services or functions offered by FCSIT DR system. They include integrity, flexibility, availability, reliability, capacity, data currency, data retention and security on the development process and standards [8].

### 2.4.1 Product Requirements

Requirements which specify that the delivered product must behave in a particular way e.g. execution speed, reliability, etc.

#### 2.4.1.1 Integrity

Integrity is the quality of correctness, completeness, wholeness, soundness and compliance with the intention of the creator of the data. It is achieved by preventing accidental or deliberate unauthorized insertion, modification or destruction of data in its database. Data integrity is one of the six fundamental components of information security.

#### 2.4.1.2 Flexibility

The programmer and designer built with regard to the customer requirements. The library needs to solve the dissertation tracking problem and controlling the interaction process of downloading, uploading and searching of its available materials.

#### 2.4.1.3 Availability

There are two concepts for availability, which is; hours of operation and reliability. The first refers to what time a dissertation repository system, which will be available for production. The second refers to its availability during the stated hours of operation of our system. I supplemented this post with the following one, extreme availability and reliability.

#### 2.4.1.4 Reliability

As the information in the dissertation repository system is safe and unchangeable without the administrator's knowledge, therefore this system is safe and can be dependent on.

#### 2.4.1.5 Capacity

Capacity deals with the projected load that our dissertation repository system will handle. This includes its growth and the timing around when heavy load conditions will occur. As dissertations move towards a more study oriented approach it becomes very important to be able to understand capacity. 8.1.7 Data Currency
Data currency is about how up-to-date our information needs to be. Data availability depends largely on the digitisation process of old thesis and dissertations from year 1988 to 2006 and the database for our system will keep up to date any new thesis and sertations reporting by the new graduates to our dissertation repository system.

#### 2.4.1.6 Data Retention

Dissertation depository system users need addresses to store and dispose information or thesis. FCSIT DR has a legislation surrounding the acquisition and disposal of information.

#### 2.4.1.7 Usability

Dissertation repository system is a well structured system with user manuals, informative errors messages, help facilities and consistent interfaces enhance usability.

#### 2.4.1.8 Space

Space refers to processor speed, memory, disk space, network bandwidth and other space limits which is suitable for our system.

### 2.4.2 Organizational Requirements

Requirements which are a consequence of organisational policies and procedures e.g. process standards used, implementation requirements, etc.8.2.1 Upgradeability
The DRS uses a service-oriented architecture approach, this system will handle web site, and calculates/ counts dissertations and then sends information to the fulfillment group. In case the rules change, we can build or purchase a replacement component to handle the changes/ calculations. The components of the DRS are loosely-coupled in a service-oriented architecture; therefore changes to one component will be transparent to other components (as long as the service itself functions the same way.). This shows the upgradeability of the DRS is high.

#### 2.4.2.1 Maintainability

The implementation process of dissertation repository system contains system software preparation and transition activities, such as the conception and creation of the maintenance plan, the preparation for handling problems identified during our system development, and the follow-up on product configuration management.

### 2.4.3 External Requirements

Requirements which arise from factors which are external to the system and its development process e.g. interoperability requirements, legislative requirements, etc [9].

#### 2.4.3.1 Security

This requirement describes on how our system handles customer privacy as well as user privileges. Security is about ensuring the safety of system. It means only authorized users can access data at permitted times. If a user wants to access the system, the user will need to have a user name and a password.

#### 2.4.3.2 Legislative

The dissertation repository system must meet the legislative requirements. This is an important task because a system or software must have licence agreement to legitimise it.

## 3. HIGH - FIDELITY PROTOTYPE

High - fidelity prototype a prototype that is quite close to the final product, with lots of detail and functionality. From a user testing point of view, a high-fidelity prototype is close enough to a final product to be able to examine usability questions in detail and make strong conclusions about how behavior will relate to use of the final product.

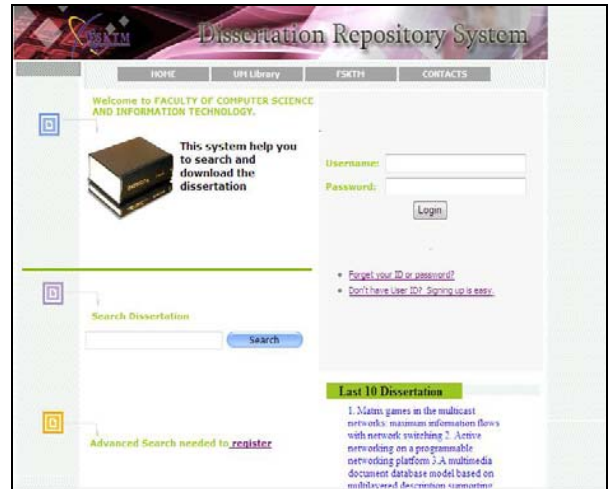

Fig 3. Main Page (Index)

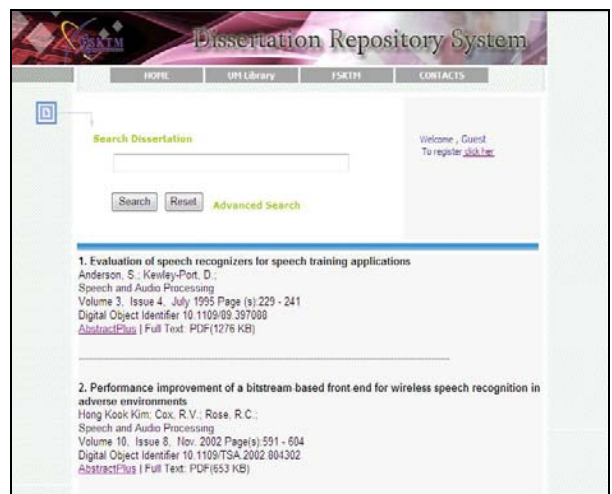

Fig 4. Guest Search Result

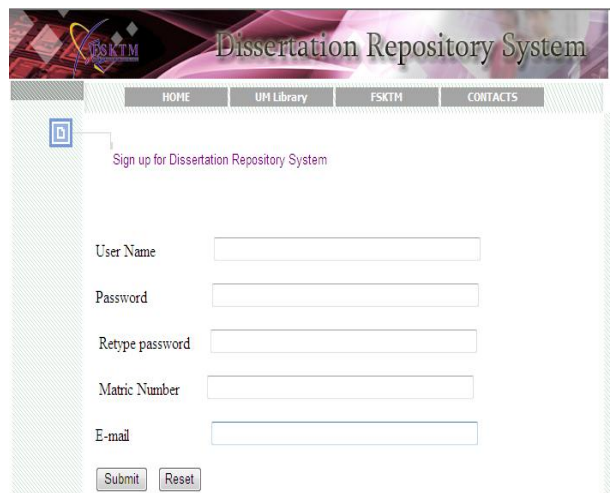

Fig 5. Registration Page

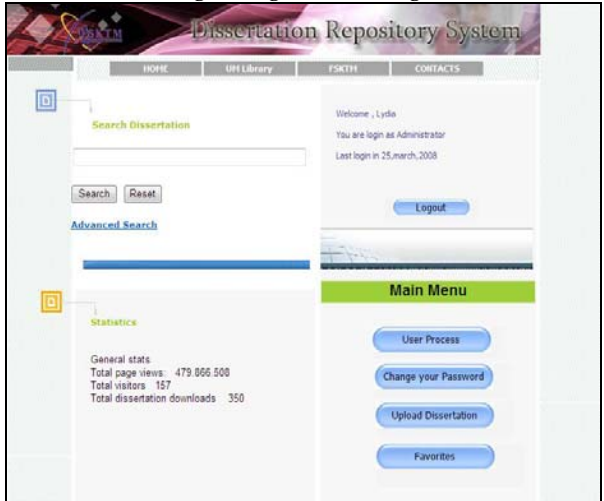
Fig 6. Administrator Home Page

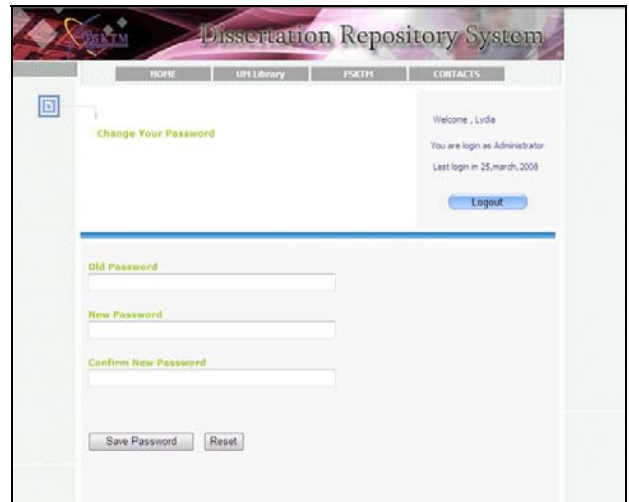
Fig 9. Administrator Change Password Page

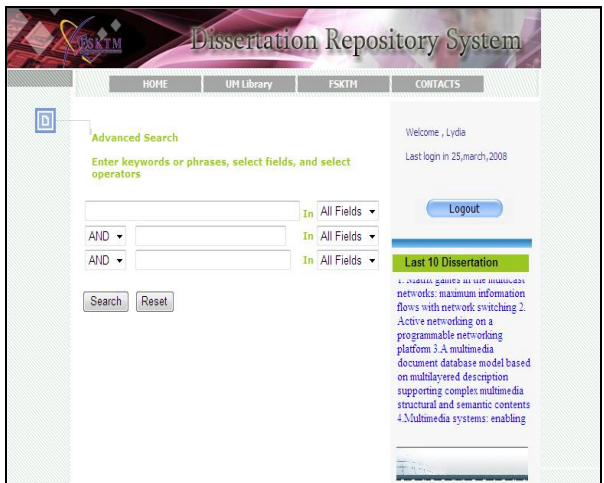
Fig 7. Administrator Advanced Search Dissertation

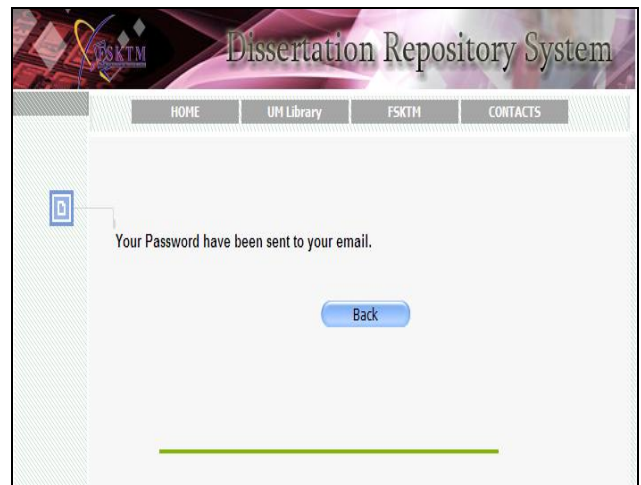
Fig 10. Administrator Password Changed Page

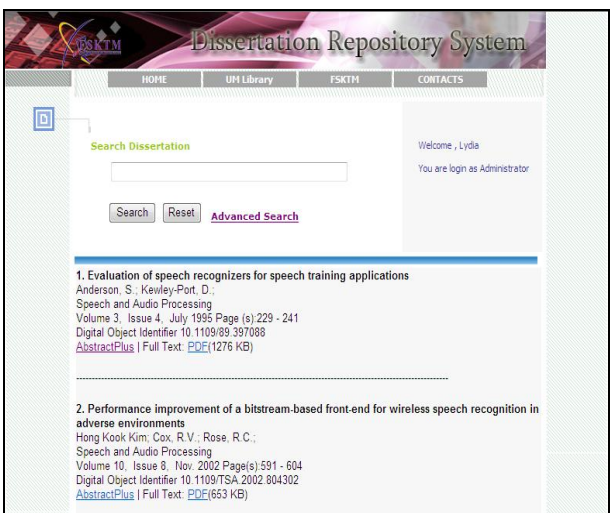
Fig 8 Administrator Search Dissertation Result

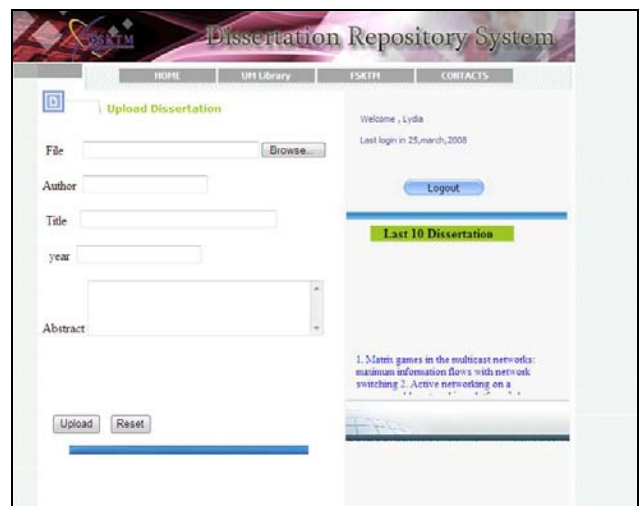
Fig 11. Administrator Upload Dissertation

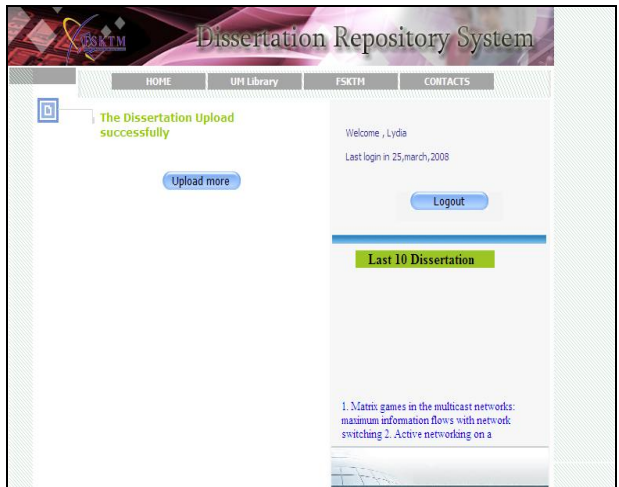
Fig 12. Administrator Upload Dissertation Successful

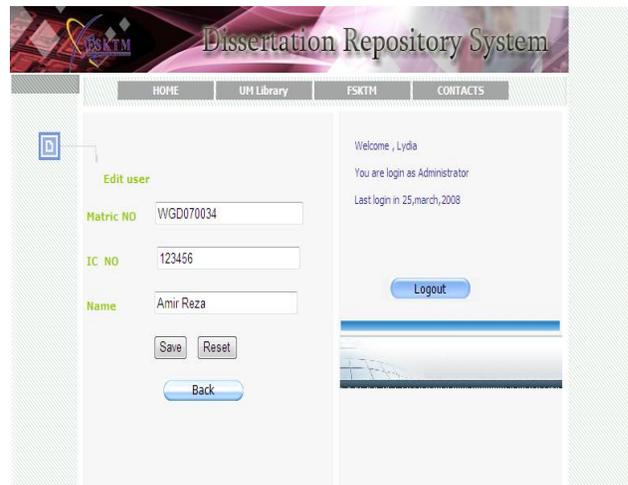
Fig 15. Administrator Edit Member

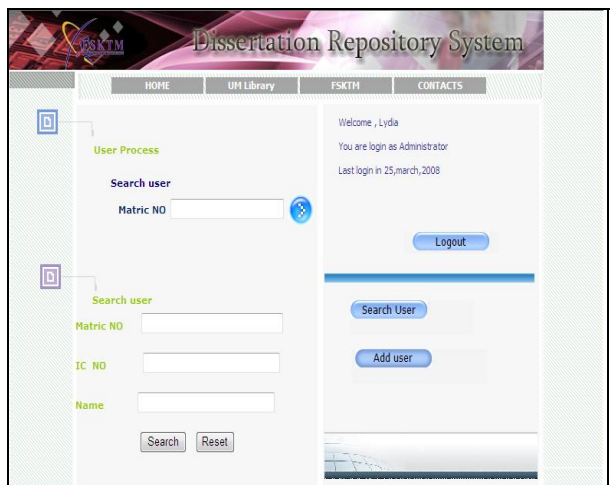
Fig 13. Administrator Member Processes

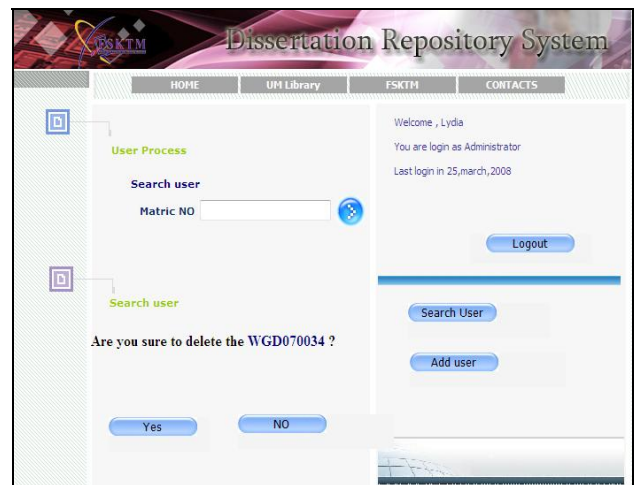
Fig 16. Administrator Delete Member

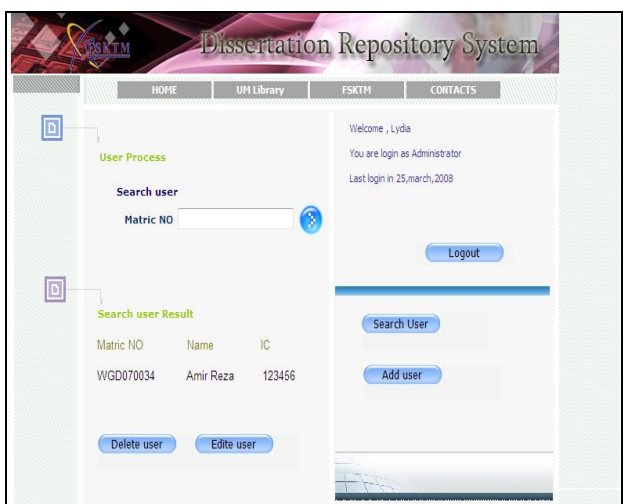
Fig 14.  Administrator Search Member

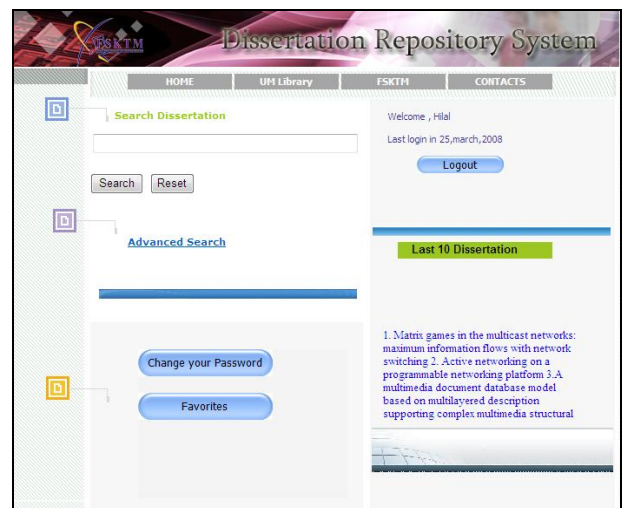
Fig 17. Member Home Page

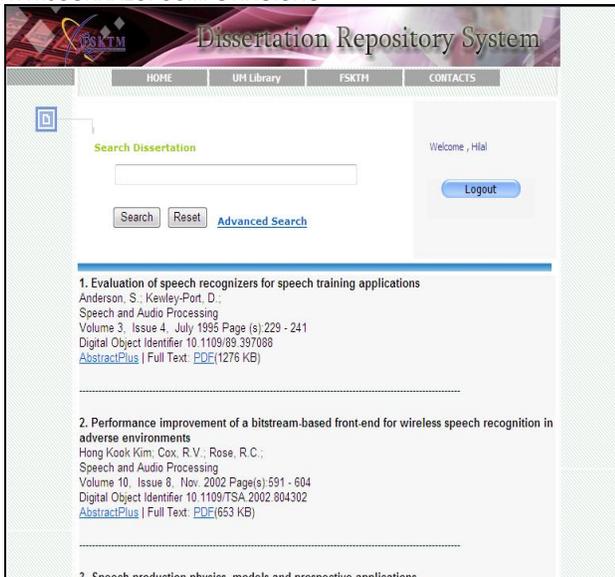

Fig 18. Member Search Result Page

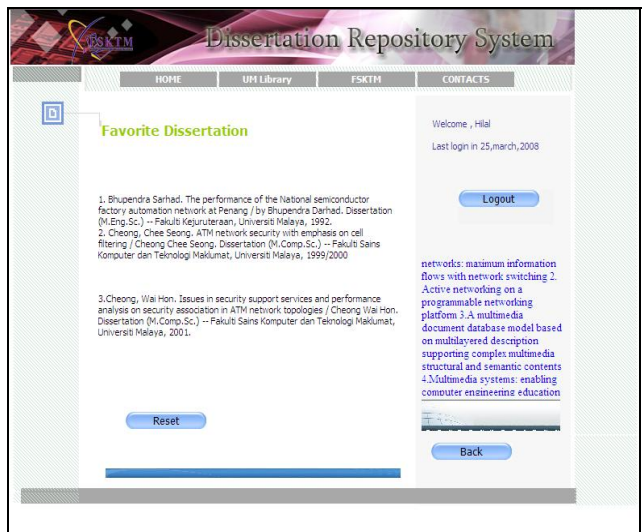

Fig 19. Member Favorite Page

## 4. CONCLUSION

In this paper, we have presented Dissertations Repository System Using Context Module. As you have seen, the system can help the students to retrive and search about specific dissertation rapidly. Also, there are Functional Requirement and Non-Functional Requirements as we have mentioned above.

## ACKNOWLEDGEMENT

This research was fully supported by "King Saud University", Riyadh, Saudi Arabia. The author would like to acknowledge all workers involved in this project that had given their support in many ways, aslo he would like to thank in advance Dr. Musaed AL-Jrrah, Dr. Abdullah Alsbail, Dr. Abdullah Alsbait. Dr.Khalid Alhazmi , Dr. Ali Abdullah Al-Afnan, Dr.Ibrahim Al-Dubaian and all the staff in king Saud University especially in Applied Medical Science In "Al-Majmah" for thier unlimited support, without thier notes and suggestions this research would not be appear.

**Ali K.Hmood -** he is master student in the Department of Software Engineering / Faculty of Computer Science and Information Technology/University of Malaya /Kuala Lumpur/Malaysia, He has contribution for many papers at international conferences and journals

**Mussab alaa Zaidan -** he is master student in the Department of Information Technology / Faculty of Computer Science and Information Technology / University of Malaya/ Department /Kuala Lumpur/Malaysia, He has contribution for many papers at international conferences and journals.

**Hamdan Al-Anazi**: He has obtained his bachelor degree from "King Saud University", Riyadh, Saudi Arabia. He worked as a lecturer at Health College in the Ministry of Health in Saudi Arabia, and then he worked as a lecturer at King Saud University in the computer department. Currently he is Master candidate at faculty of Computer Science & Information Technology at University of Malaya in Kuala Lumpur, Malaysia. His research interest on Information Security, cryptography, steganography, Medical Applications, and digital watermarking, He has contributed to many papers some of them still under reviewer.

**Rami Alnaqeib -** he is master student in the Department of Information Technology / Faculty of Computer Science and Information Technology/University of Malaya / Kuala Lumpur/Malaysia, He has contribution for many papers at international conferences and journals.

**Yahya Al-Nabhani -** he is master student in the Department of Computer System and Technology / Faculty of Computer Science and Information Technology/University of Malaya /Kuala Lumpur/Malaysia, He has contribution for many papers at international conferences and journals.